\documentclass[11pt]{article}   	
\pdfoutput=1
\usepackage[utf8]{inputenc}
\usepackage{jheppub}

\usepackage{graphicx}				
\usepackage{amssymb}
\usepackage{amsmath}
\usepackage{bm}
\usepackage{slashed}
\usepackage{epigraph}
\usepackage{feynmf}

 \usepackage{natbib}

\newcommand{\beq}{\begin{eqnarray}}
\newcommand{\eeq}{\end{eqnarray}}

\newcommand{\avg}[1]{\left\langle#1\right\rangle}

\newcommand{\braket}[3]{\left\langle#1\left| #2\right| #3\right\rangle}

 \newcommand{\be}{\begin{equation}}
 \newcommand{\ee}{\end{equation}}
 \newcommand{\sign}{\text{sign}}
\newcommand{\R}{\mathbb R}
\renewcommand{\S}{\mathbb S}

\title{The curious incident of multi-instantons and 
the necessity of Lefschetz  thimbles
}
\author[a]{Alireza Behtash,}
\author[b]{Erich Poppitz,}
\author[a]{Tin Sulejmanpasic,} 
\author[a]{Mithat \"Unsal}

\affiliation[a]{Department of Physics, North Carolina State University, Raleigh, NC 27695, USA}
\affiliation[b]{Department of Physics,   University of Toronto, 
Toronto, ON M5S 1A7, Canada}

\emailAdd{abehtas@ncsu.edu}
\emailAdd{poppitz@physics.utoronto.ca}
\emailAdd{tsulejm@ncsu.edu}
\emailAdd{unsal.mithat@gmail.com}

\abstract
{We show that compatibility of supersymmetry with exact semi-classics demands 
that in calculating multi-instanton amplitudes, the ``separation" quasi-zeromode must be complexified and the integration cycles must be found by using complex gradient flow (or Picard-Lefschetz equations.) 
As a  non-trivial application,  we study $\mathcal N=2$ extended supersymmetric quantum mechanics.  
 Even though in this case supersymmetry is unbroken, the instanton--anti-instanton amplitude (naively calculated) seems to contribute  to  the ground state energy.   We show, however, that the instanton--anti-instanton event consists of two parts: a fermion-correlated and a scalar-correlated event. Although both of these contributions are naively of the same sign and the latter is superficially higher order in the perturbative coupling,  we show that the two contributions  exactly cancel  when they are evaluated on Lefschetz thimbles due to their  relative  Hidden Topological Angles (HTAs). This gives strong evidence that the semi-classical expansion using  Lefschetz thimbles is  not only a meaningful prescription for higher order semi-classics, but a necessary one. 
 This deduction  seems to be universal and applicable to both supersymmetric and non-supersymmetric theories. In conclusion we speculate that similar conspiracies are responsible for the non-formation of certain molecular contributions in theories where instantons have more than two fermionic zeromodes and do not contribute to the superpotential.

}
\begin{document}
\maketitle
 
\setlength{\epigraphwidth}{0.825\textwidth}
\epigraph{
   Gregory: ``Is there any other point to which you would wish to draw my attention?"\\
    Holmes: ``To the curious incident of the dog in the night-time."\\
       Gregory: ``The dog did nothing in the night-time."\\
    Holmes: ``That was the curious incident."
}{\vspace{.2cm}\emph{Silver Blaze} by Arthur Conan Doyle}

\section{Introduction}

Instantons---the prototypical semiclassical objects---have been of interest in quantum field theory and quantum mechanics  for a long time. They play   
instrumental roles in  virtually every field theory with nontrivial infrared (IR) physics.    Whenever the quantum theory under consideration satisfies  semi-classical calculability,  instantons provide the key to understanding the long distance physics and explaining phenomena such as mass-gap generation in non-supersymmetric  QFTs  \cite{Polyakov:1976fu,Unsal:2008ch}. 
They also provide the  origin of  non-perturbatively induced superpotentials in  many  supersymmetric QFTs, see e.g.  lecture notes in \cite{Deligne:1999qp}.  Instantons also play a role in phenomenological models of  chiral symmetry breaking in QCD,  see \cite{Schafer:1996wv}.

A major obstacle that appears already at weak coupling is that evaluating multi-instanton contributions to observables is not only a formidable task, but no precise rationale  exists for this procedure. The trouble comes from the fact that instanton--anti-instanton configurations belong to the perturbative vacuum, and naive integration over their   separation mixes the perturbative contribution with the non-peturbative one. On the other hand,  the desire to incorporate  multi-instanton configurations systematically  is not aimed at finding sub-leading corrections to the instanton effects, which would be a relatively dull task. Rather, it is inspired by two observations regarding multi-instantons:  { 1.)} There are qualitatively new effects arising from them,  e.g. the  vacuum energy in supersymmetric QM \cite{Balitsky:1985in} and mass gap in QCD(adj) on small $\R^3\times \S^1$ \cite{Unsal:2007jx}.  { 2.)} The realization that they play a crucial role in the resurgent transseries expansion in QM and QFT, where 
multi-instanton effects can indirectly  be calculated  (in the case of QM) via exact quantization  conditions  \cite{ZinnJustin:2002ru} and the uniform WKB approach \cite{Dunne:2014bca}, establishing remarkable connection between perturbative and nonperturbative sectors. This connection was explicitly checked for the double well potential \cite{Escobar-Ruiz:2015nsa} and the sine-Gordon potential \cite{Escobar-Ruiz:2015rfa} to three loops providing a direct confirmation of  \cite{Dunne:2014bca}, while the  resurgent structure of sine-Gordon potential  was checked to match the uniform WKB \cite{Misumi:2015dua}. 

Historically crucial progress in understanding the case of quantum mechanics (QM) was made by Bogomolny   \cite{Bogomolny:1980ur} and Zinn-Justin \cite{ZinnJustin:1981dx,ZinnJustin:1982td} long ago. They proposed a prescription, which is   called the  ``BZJ-prescription" in  \cite{Argyres:2012ka,Dunne:2012ae}, 
 for evaluating instanton--anti-instanton contributions,  incorporating  an analytic continuation in the coupling. 
  Soon after, Balitsky and Yung  argued in \cite{Balitsky:1985in} that  a certain complex multi-instanton quasi-solution  should be taken into account to explain the positive sign of the energy in supersymmetric QM with spontaneously broken supersymmetry.    
But there was little hope of extending these methods to quantum field theory (QFT). 
Recently the BZJ-prescription  was successfully applied \cite{Argyres:2012ka,Poppitz:2012sw,Poppitz:2012nz} to the case of QCD with adjoint matter on $\R^3\times \S^1$ and non-linear sigma models on  $\R^1 \times \S^1$.    For $n_f=1$  (one adjoint Weyl fermion or ${\cal N}=1$ SYM theory), this produces the correct bosonic potential for the Polyakov loop  along with a magical  center-stabilizing  minus sign,  (A phenomenological explanation of this minus sign was given in 
 \cite{Shuryak:2013tka}.)  and for bosonic ${\mathbb CP}^{N-1}$, this procedure provides a mechanism of ambiguity cancellation in QFT, which is an essential ingredient of resurgence structure. 
This  provides crucial evidence that these ideas are applicable beyond quantum mechanics \cite{Dunne:2012ae,Dunne:2012zk,Cherman:2013yfa,Cherman:2014ofa, Misumi:2014rsa,Misumi:2014bsa,Misumi:2014jua}.  

However, the BZJ prescription is partly a black-box, and is not always fully satisfactory.  One can, by using the WKB method  in quantum mechanics, show that it produces the correct result, but there are certainly cases in which it does fail, an example of which is discussed in this work.  It would be much more useful to gain a more direct  geometric understanding  on how to treat higher order semiclassical corrections precisely. 
 
  Refs.~\cite{Behtash:2015kna, Behtash:2015, BDSSU} argued that the proper  framework to deal will multi-instanton calculus is a complex version of Morse theory, called Picard-Lefschetz theory, applied to the quasi-zero mode integrations.    
For bosonic models with instantons, this was understood in an unpublished work \cite{CDU}.      We will call the associated cycles of integrations over
 quasi-zero modes (QZM)  Lefschetz thimbles,  ${\cal J}^{\rm qzm}$. (Other applications of Picard-Lefschetz theory to 
path integrals can be found in, e.g.~\cite{Witten:2010cx,Witten:2010zr,Harlow:2011ny,Kanazawa:2014qma,Tanizaki:2015pua,Basar:2013eka, Cherman:2014ofa}.) 
Ref.~\cite{Behtash:2015kna,Behtash:2015,BDSSU} followed two complementary approaches. First, by introducing a new formalism in which configuration space is complexified,  it showed the existence of  new {\it exact} solutions governing the correct ground state properties. It also showed that 
 the corresponding complex finite action  classical solutions need not even be smooth, they can be multivalued and singular, a result surprising in itself!
 The  consistency of supersymmetry algebra and the realization of supersymmetry (broken or unbroken) is shown to be due to the interplay of certain  complex  and real saddle contributions,  see \cite{Behtash:2015kna,Behtash:2015,BDSSU} for details.    Second,  it also showed that the most salient features of the exact solutions can  be easily produced  by integrating over QZM Lefschetz  thimbles, ${\cal J}^{\rm qzm}$, between instantons and anti-instantons. Namely, the instanton--anti-instanton configuration on the thimble is  an approximation to the exact solutions mentioned above.   
 We also note that the broader context for our work is the  connection between the  (complex)  saddles of  complexified path integrals, and resurgence theory and transseries representation of path integrals. 
(Other applications of resurgence theory  in the matrix models and topological string theory context can be found in e.g.  
\cite{Marino:2012zq,Aniceto:2011nu,Aniceto:2013fka}.)

 In an ${\cal N}=1$ supersymmetric quantum mechanics, even if the theory possesses $k$ classical  harmonic minima, the Witten index is  $ |I_W|= k \;  ({\rm mod} \;2 )$.  Namely, the above mentioned instanton--anti-instanton configurations lift all possible Bose-Fermi  pairs of harmonic vacua, and one is left with either  $|I_W|= 0\; {\rm or}\; 1$ due to lifting.

In this work, we make another step in understanding the treatment of multi-instantons in semi-classics, this time in extended  $\mathcal N=2$ supersymmetric QM (four real supercharges).  In this model, it is an exact result  that all classical ground states remain quantum ground states: if such a theory  possesses $k$ classical  harmonic minima, then the Witten index is  $ |I_W|= k$.  On the other hand, the multi-instantons are present,  
but they just do  ``{\it nothing}." This paper is about this ``{\it nothing}," which, in turn,  provides  new insights into an exact version of the semi-classical method. 

We shall see that a subtle cancellation of the instanton--anti-instanton  contribution to the vacuum energy occurs.
We show that to leading order in the semiclassical expansion there are two contributions to the correlated instanton--anti-instanton event: $i.)$ a fermion-correlated instanton--anti-instanton event and $ii.)$ a contribution lifting zero modes via the Yukawa coupling and a scalar exchange instead. While the latter contribution is formally higher order in the coupling, it exchanges only one massive scalar, while the fermion-correlated event exchanges two massive fermions. The suppression factors of the two events at large instanton--anti-instanton separation $\tau$ are then of order $e^{-2\omega \tau}$ and $e^{-\omega\tau}$, respectively, where $\omega$ is the harmonic oscillator frequency. As we shall see explicitly in the case of the double well potential, the integration over $\tau$, when defined as an integration on appropriate steepest descent paths, or Lefshetz thimbles, leads to \emph{exact} cancellation between these two contributions. Thus, the instanton--anti-instanton contribution to the potential vanishes, consistent with the unbroken supersymmetry---but only if the integration over the quasi-zeromode is done on the complex steepest-descent path. This suggests that the integration over Lefshetz thimbles is a crucial ingredient in extending semi-classics beyond the leading order. We should emphasize that this cancellation is  different from that of $\mathcal N=1$ QM with unbroken SUSY where the cancellation is between a real and a complex saddle \cite{Behtash:2015kna,Behtash:2015,BDSSU}. In contrast in $\mathcal N=2$ QM the cancellation is between two \emph{complex} saddles, or better yet, between two thimbles associated with the two complex saddles.  The cancellation in both case arise due to an $e^{i  \pi} \in {\mathbb Z_2}$ worth of hidden topological angle phase difference between the two distinct thimbles.

 This paper is organized as follows. The reader interested   in the main features of the result  will be satisfied with reading Section~\ref{sec:formulation} only. There, we present the model and sketch the cancellation of the instanton--anti-instanton contribution to the vacuum energy described above, stressing the importance of integration over Lefshetz thimbles. Section~\ref{sec:details} gives significantly more detail on the derivation of the main result.  
 We conclude in Section~\ref{sec:discuss}.

\section{Basics of $\mathcal N=2$ supersymmetric quantum mechanics}\label{sec:formulation} 

We consider $\mathcal N=2$ supersymmetric (SUSY) quantum mechanics (QM). It is obtained by dimensional reduction of the 4D Wess-Zumino model of a single chiral superfield $z$ and arbitrary superpotential $W(z)$ down to quantum mechanics. The Euclidean Lagrangian is
\be\label{eq:lag}
g\mathcal {L}_E=|\dot z(t)|^2+|W'(z)|^2+\begin{pmatrix}\bar\chi_1&\chi_2\end{pmatrix}\left(-\partial_t+\begin{pmatrix}0&\overline{W''(z)}\\
{W''(z)}&0\end{pmatrix}\right)\begin{pmatrix}\chi_1\\\bar\chi_2\end{pmatrix}\;,
\ee
where
\be
z(t)=x(t)+iy(t)
\ee
 is the complex coordinate of the particle and $\chi_{1,2}(t),\bar\chi_{1,2}(t)$ are Grassmann-valued coordinates of the particle.\footnote{As opposed to field theory, the Grassmann fields do not represent separate particles, but instead endow a 2D quantum particle at $(x,y)$ with a spin degree of freedom, which is spin $\frac{1}{2} \otimes \frac{1}{2} $ because of ${\cal N}=2$ structure.
} 
Further below,  we specialize to the case of the double-well potential with $k=2$, and   $W(z)={1\over 3} z^3 - z a^2$, taking $a$ real  without loss of generality. The  frequency around the minima of the bosonic potential, $z_\pm = \pm a$, is  $\omega = 2a$. Upon rescaling, it is seen that anharmonic terms are multiplied by $\sqrt{g}$  of dimension $\omega^{3 \over 2}$. In this paper, we focus on the semiclassical limit ${g \ll \omega^3}$.
The  action is invariant under the SUSY transformation
\begin{subequations}
\label{susyeqns}
\begin{align}
&\delta z=\sqrt2(\epsilon_2\chi_1-\epsilon_1\chi_2)\;,&&\delta \bar z=\sqrt2({\bar\epsilon_1\bar\chi_2-\bar\epsilon_2\bar \chi_1})\;, \\
&\delta\chi_1=\sqrt{2}(-\dot z\bar\epsilon_2-\overline{W'}\epsilon_1)\;,&&\delta\bar\chi_1=\sqrt{2}(\dot {\bar z}\epsilon_2-{W'}\bar\epsilon_1)\;,\\
&\delta\chi_2=\sqrt{2}(\dot z\bar\epsilon_1-\overline{W'}\epsilon_2)\;,&&\delta\bar\chi_2=\sqrt{2}(-\dot {\bar{z}}\epsilon_1-{W'}\bar\epsilon_2)\;.&  
\end{align}
\end{subequations}

The critical points of the superpotential, assumed nondegenerate, $W'(z_i)=0$, $z_i$, $i=1,\dots k$ ($k=2$ for our cubic $W$) are the classical minima of  the bosonic potential $|W'(z)|^2$. It has been known for a long time that all classical ground states remain quantum-mechanical ground states \cite{Jaffe:1987nx} (see also ~Ch.~10 in \cite{Hori:2003ic}).   To quickly review the argument, recall that the Witten index is invariant under continuous deformations of the potential, in particular under rescaling of the superpotential $W\rightarrow \sigma W$. Taking first $\sigma\rightarrow\infty$, the theory is well approximated by $k$ distinct SUSY quantum harmonic oscillators.  
In a harmonic approximation,  quantizing the system on the left and the right well, we obtain 
\begin{align}
H_{L, R} = |\Pi_z|^2 + (\pm 2a)^2 |z|^2 + (\pm 2a) (a_1^{\dagger} a_2^{\dagger} + a_1 a_2 ) \;,
\end{align}
where $a_i^{\dagger},  a_i$ $(i=1,2)$ are fermion creation/annihilation operators. 
The harmonic ground states on the left well and right well are given by 
\begin{align}
| L, {\bf 0} \rangle_b \otimes  \left(  |\uparrow \uparrow \rangle   +  |\downarrow \downarrow \rangle  \right), \qquad  | R, {\bf 0} \rangle_b \otimes  \left(  |\uparrow \uparrow \rangle   -  |\downarrow \downarrow \rangle  \right), 
\end{align}
both of which are bosonic, and there are no fermionic partners. Fermionic states involving  $|\uparrow \downarrow \rangle, |\downarrow \uparrow \rangle$ are 
excited states. Since in a supersymmetric theory, all positive energy states are 
Bose/Fermi paired by supersymmetry, and  states can only ascend/descend in  Bose/Fermi pairs, the two bosonic ground states can never be lifted.  
Thus the Witten index is nonzero $(I_W=2)$  and supersymmetry is unbroken. Further, 
none of the classical ground states can be lifted by perturbative or nonperturbative (instanton or multi-instanton) effects, thus they all remain true ground states of the full quantum theory.  

\vspace{0.5cm}
{\bf Difference between $ {\cal N}=1$ and $ {\cal N}=2$ QM, and a puzzle:}
Note the  sharp contrast between 
$ {\cal N}=1$ supersymmetry, with real superpotential $W(x)$  and the $ {\cal N}=2$  theory with holomorphic superpotential $W(z)$, e.g. 
\begin{align} 
W(x) =\prod_{i=1}^{k+1} (x-x_i)    \qquad {\rm vs.}  \qquad  W(z) =\prod_{i=1}^{k+1} (z-z_i) 
\end{align}
  In the  $ {\cal N}=1$  case,   the  harmonic zero energy ground states in any two consecutive  harmonic wells are always alternating, if one is bosonic, the other  is strictly  fermionic.   Consequently,  since a Bose-Fermi paired zero energy state can  happily move up simultaneously, 
 in $ {\cal N}=1$ supersymmetry, lifting  happens generically.   In the  $ {\cal N}=2$, this is never the case. All harmonic grounds states are either fermionic or bosonic, and hence, the zero energy levels can never be lifted.   
 Consequently, if the number of critical points    is $k$, the Witten index is,  
 \begin{align} 
& |I_W|= k \;  ({\rm mod} \;2 )  \qquad \;\; {\cal N}=1, \cr
&  |I_W|= k \;  \qquad \qquad \qquad {\cal N}=2.
\end{align}
The lifting of the harmonic zero energy states  cannot happen perturbatively, but may  happen non-perturbatively. In the  ${\cal N}=1$ case, this  provides the
$k$ low-lying   states with  energies $\sim e^{- 2S_0 /g}$ (where $S_0/g$ is the instanton action) or zero.   Strictly, the energies of low lying levels arise from a multi-instanton effect, and not an instanton.  On the other hand, in the ${\cal N}=2$ case,   instantons and  multi-instantons  seem to do {\it nothing}.  This is the curious incident that we would like to understand by semi-classical methods, instead of relying on supersymmetry. Our  hope  is to learn  something important about the nature of the semi-classical method, which is more widely applicable than the supersymmetric techniques.

\subsection{The curious incident of instantons in ${\cal N}=2$ QM, and the necessity of thimbles}\label{sec:formulation-2}
Although  the non-lifting of the zero energy grounds states in  ${\cal N}=2$ QM   is well known,  it  may at first appear strange to someone not familiar with the constraints of (extended) supersymmetry.  Tunnelling events between vacua should be present on general grounds and  are expected to lift the vacuum degeneracy in non-supersymmetric theories  by level splitting,    and by simultaneously  lifting  Bose-fermi paired harmonic minima in  ${\cal N}=1$ QM.   In both ${\cal N}=1$ and ${\cal N}=2$,  if this  lifting is to happen, it cannot be facilitated by a single instanton due to fermion zeromodes.  
Thus, the leading-order semiclassical contribution is an instanton--anti-instanton molecular event, similar to the ones considered long ago \cite{Bogomolny:1980ur,ZinnJustin:1981dx,ZinnJustin:1982td}. 

\begin{figure}[htbp] 
   \centering
   \includegraphics[width=.7\textwidth]{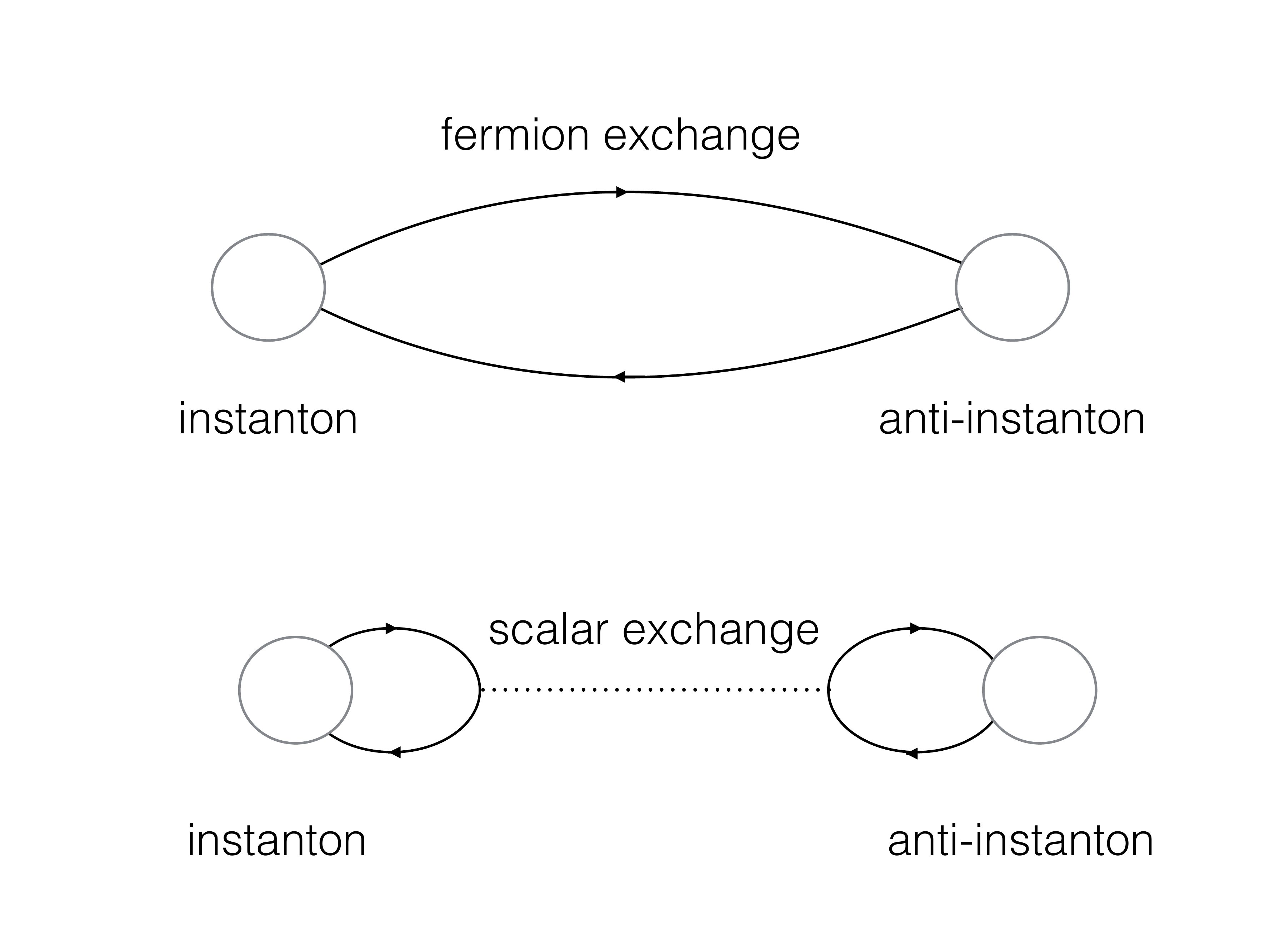} 
   \caption{Top: a fermion-correlated $I\bar I$ event, contributing the first term in Eq.~(\ref{eq:groundstate}). Bottom: a scalar-correlated $I\bar I$ event, contributing the second term in Eq.~(\ref{eq:groundstate}). The two contributions are proportional  to different powers of the perturbative  ($g \ll \omega^3$) coupling $g$.  In QM, the diagrams  are  intended to schematically represent the lifting of fermion zero modes by the two mechanisms. In QFT,   one can associate the (anti-)instanton vertices  with effective  't Hooft interactions and the lines connecting them to free (away from the  instanton cores) scalar and fermion propagators.}
   \label{fig:diagrams}
\end{figure}

In order for the   the instanton--anti-instanton molecular event to contribute to the vacuum energy, the fermion zeromodes have to be lifted. One way this lifting can arise can be thought of   as  due the exchange of the fermionic zeromodes, as in the top diagram on  Fig.~\ref{fig:diagrams}.
Another way to lift the fermion zero modes  is due to background scalar fluctuations of the $y(t)$ field (fluctuations of $x(t)$ do not contribute, see Section \ref{sec:details}), which couples to the (anti-)instanton via the Yukawa coupling, as in the bottom diagram on Fig.~\ref{fig:diagrams}.  
Naively, the Yukawa vertex coupling the fermions to the scalar  makes this contribution subleading in the small coupling $\sqrt{g}$. 

In Section \ref{sec:details}, we compute these two contributions and show that the two kinds of correlated events  contribute to the ground state energy in a following manner
\be\label{eq:groundstate}
E_0\propto -e^{-2S_0}\int d\tau\;e^{\frac{4\omega^3}{g}e^{-\omega\tau}}\left(4\omega^3 e^{-2\omega\tau}+g e^{-\omega\tau}\right)  
\equiv  
- e^{-2S_0}\int d\tau \; (  e^{ -V_1 (\tau) }  +  e^{ -V_2 (\tau) } )
\;.
\ee
At this stage, $\tau$ is the instanton--anti-instanton separation, $\omega=2a$, and $S_0 = \frac{8 a^3}{3 g} = \frac{\omega^3}{3 g}$ is the action of a single instanton.  The $e^{\frac{4\omega^3}{g} e^{-\omega\tau}}$ factor in the integrand is the $I$-$\bar I$ long-distance attraction and the two factors in the brackets are the fermion-correlated, $\sim e^{-2 \omega \tau}$,  and scalar-correlated, $\sim e^{- \omega \tau}$, contributions.
Naively, the integral over the separation in (\ref{eq:groundstate}) is to be taken from $\tau=0$ to $\tau=\infty$. It seems impossible that  $E_0$ in (\ref{eq:groundstate}) can ever vanish,  as the integrand is strictly positive for any $\tau \ge 0$. As it stands, this is in contradiction with the constraints of supersymmetry,  and more disastrously,  with the supersymmetry algebra which demands that energy  is positive semi-definite. But the story is more subtle, and one with happy ending.

 \begin{figure}[htbp] 
   \centering
   \includegraphics[width=.7\textwidth]{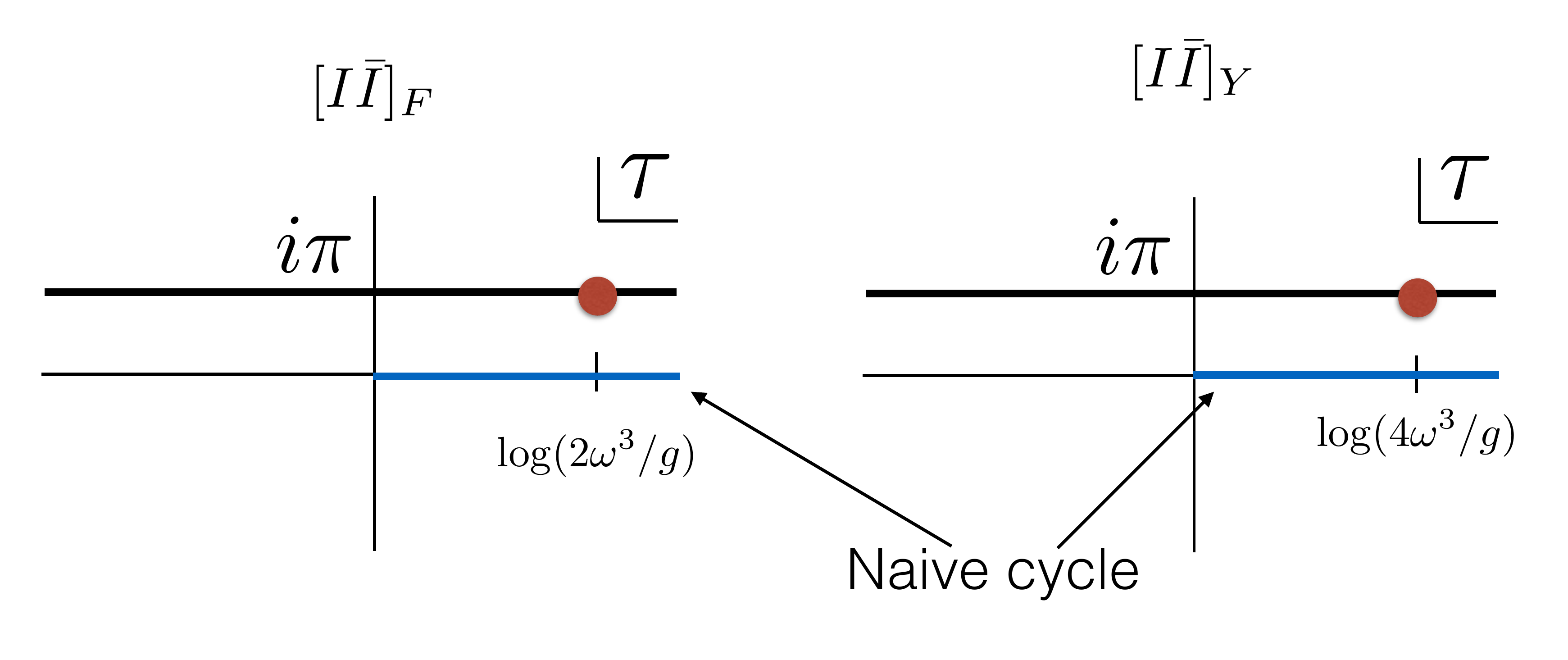} 
   \caption{The steepest descent cycles for the fermion-correlated channel vs. scalar correlated channels. The blue cycle is the naive cycle in which the separation between the instanton and anti-instanton is interpreted as real. A result compatible with supersymmetry only comes about if we use the critical point cycles. 
}
   \label{fig:cycles}
\end{figure}

As argued in \cite{Balitsky:1985in} and  formalized more recently in \cite{Behtash:2015,BDSSU, CDU} in the context of  resurgence and Picard-Lefschetz theory, the integral should be thought of as an integral  in the complex $\tau$ plane. Since $\tau$ corresponds to some field direction, its complexification is to be thought of as the complexification of the original fields, which are to be treated by   complex gradient flow (Picard-Lefschetz) equations.    Of course, the full complexified field space is infinite dimensional, and in principle, we have to work in the context of the  Picard-Lefschetz equations for the full theory. 
However,  in the background of multi-instanton  saddles, as concrete evidence is provided in  \cite{Behtash:2015kna,BDSSU, CDU},  
 this space usually factorizes into finite dimensional zero and quasi-zero modes directions and infinite dimensional gaussian modes: 
 \begin{align}
{\cal J}^{\rm full} ={\cal J}^{\rm Gaussian} \times {\cal J}^{\rm zm} \times  {\cal J}^{\rm qzm}  ~.
 \end{align} 
In the determination of the correlated instanton--anti-instanton contribution to ground state energy, the most important  subcomponent of the thimble ${\cal J}^{\rm full}$, which governs some of the salient features of the multi-instanton configuration, is  ${\cal J}^{\rm qzm}$.   
This reduces a formidable task of treating an infinite dimensional path integral  to that of treating an interesting  finite (in this case one-) dimensional integral by Picard-Lefschetz theory and  a much less interesting  infinite dimensional Gaussian integration. 

Accepting Eq.~\eqref{eq:groundstate} for the moment (it is one of our main results and will be carefully derived in the  Section \ref{sec:details}), we define the following integrals
\begin{subequations}\label{eq:J1J2thimble}
\begin{align}
&I_1=\int_{{\cal J}_1} d\tau\; e^{\frac{4\omega^3}{g}e^{-\omega\tau}-2\omega\tau}\;,\\
&I_2=\int_{{\cal J}_2} d\tau\; e^{\frac{4\omega^3}{g}e^{-\omega\tau}-\omega\tau}\;,
\end{align}
\end{subequations}
and identify  
\begin{align}
{\cal J}^{\rm qzm} =   {\cal J}_1 +  {\cal J}_2\; .
\label{qzm-thimble}
\end{align} 
The saddle points of the exponents in the complex $\tau$ plane are
\begin{align}
\label{saddletau}
&\omega\tau_1=i \pi + \log{2\omega^3 \over g}\;,\\
&\omega\tau_2=i \pi + \log{4 \omega^3 \over g}\;,
\end{align}
where the index $1,2$ corresponds to integrals $I_{1,2}$.\footnote{The exponent has other critical points, but since the integrand only depends on $e^{-\omega\tau}$, the values of $\tau$ are equivalent up to a $2\pi i/\omega$ shift. There are, however, two critical points which are not a priori equivalent and  differ by having Im$(\omega \tau)=\pm \pi$. Which saddle point is selected cannot be determined for real $g$. Instead $g$ should be defined as having a small imaginary part which will be sent to zero at the end of the computation. In the present case the final result will not depend on whether we selected $\text{Im} g>0$ or $\text{Im} g<0$ and which saddle point we choose to evaluate the quasi-zeromode integral. In general, for non-supersymmetric theories, this will not be the case and will cause an inherent ambiguity in semiclassical computations. In these theories, however, the ambiguity will be cancelled \emph{exactly} by the ambiguity of the perturbation theory which is caused by its non-Borel summability. The two ambiguities shall always cancel exactly leaving an unambiguous and real result for real observables. This is one of the essential features of the resurgent expansion.} The integrals are then evaluated on the steepest-descent paths, 
satisfying complex gradient flow equations: 
\begin{align}
\frac{\partial \tau}{\partial u} =   \frac{ \partial \overline V_i (\bar \tau) }{\partial \bar  \tau}  \;,
\end{align}
where $u$ is gradient flow time, and $u=-\infty$ is the critical point of  $V_i (\tau)$. 
Equivalently,  due to the one-dimensional nature of the present problem,  this cycle corresponds to the stationary phase cycle: 
\begin{align}
\text{Im }V_i (\tau) = \text{Im }V_i (\tau_i), \;\;\; {\rm i.e.}  \;\;\; 
\text{Im }( \omega\tau)=\pi
\end{align}
along the path.
It is easy to see that in both cases this corresponds to integrating on the line parallel to the real axis and shifted by $i\pi/\omega$, i.e. $\tau\in(-\infty+i\pi/\omega,\infty+i \pi/\omega)$.  This yields
\begin{align}
&I_1= \frac{g^2}{16\omega^7}\;,\\
&I_2= - \frac{g}{4\omega^4}=  \frac{4\omega^3}{g} ( e^{i \pi} I_1)\;,
\end{align}
where  $e^{i \pi}$ is the relative phase between the two thimbles, ${\cal J}_1$ and  ${\cal J}_2$---an example of a hidden topological angle \cite{Behtash:2015kna}.  Therefore,  the vacuum energy (\ref{eq:groundstate}) vanishes:
\be
E_0\propto 4\omega^3I_1+g I_2= 4\omega^3(1 + e^{i \pi}) I_1=  0\;.
\ee

Remarkably, the two contributions not only have the opposite sign, but are of the same order in $g$ and cancel exactly! How did this happen? Crucial to the cancellation was the exponential suppression $e^{-2\omega\tau}$ in the case of fermion-correlated event and $e^{-\omega\tau}$ in the case of scalar-correlated event. The critical points of both integrals are at  Re$(\tau_{1,2}\omega)\propto -\log g$. However the integrand at the critical point of $I_1$ and $I_2$ integrals contain $e^{-2\omega\tau_1}\propto g^2$ and $e^{-\omega\tau_2}\propto g$, so that although $I_1$ started initially as lower order in $g$, the exponential suppression due to fermion exchange forced the integral $I_1$  to contain an extra factor $g$ compared to the integral  $I_2$. 

We find this incredible conspiracy   nothing short of remarkable. It gives  compelling evidence that a general principle of evaluating higher order semiclassical contributions by treating their quasi-moduli via Picard-Lefschetz theory is the correct and necessary procedure.

The relative hidden topological angle among saddles  is a universal feature  seen in a broad class of supersymmetric and non-supersymmetric theories.  
In all cases studied so far, this phase difference arises from the integration over different thimbles ${\cal J}_i$ in the complex plane, whose  contributions have a relative factor of $e^{i \pi}$. For example, in ${\cal N}=1$ supersymmetric QM,  
the real cycle and complex cycle (associated with a real saddle and complex saddle) differ by  $e^{i \pi}$, while in non-supersymmetric QM with $n_f$ fermion field the relative phase is 
 $e^{i n_f \pi}$. These factors may lead to either constructive or destructive ``interference" between the contributions of different saddles. In field theory, the cleanest example is given by comparing the contributions of  the magnetic bion vs. neutral bion cycle in QCD(adj) with $n_f$ 
 flavors of fermions. There, the relative phase is $e^{i (4n_f -3) \pi}$ which, for positive integer  $n_f$, is always   $e^{i \pi}$ \cite{Poppitz:2011wy,Argyres:2012ka}.   This  overall sign is of physical significance,  and  reflects the fact that neutral  bions induce a center-stabilizing potential for any physical value of $n_f$.   In the problem considered in this paper,   it is  two distinct complex cycles (instead of one real vs. one complex)    which have a relative $e^{i \pi}$ phase.  

We will now proceed to show explicitly how the contributions $I_1$ and $I_2$ to (\ref{eq:groundstate}) arise.

\section{Computation of $\mathbf{I}$-$\mathbf{\bar I}$ contributions to the ground state energy}
\label{sec:details}

In this Section, we analyze in detail the $I\bar I$ contributions starting from the Lagrangian \eqref{eq:lag}.  Instantons are solutions of the BPS equation
\be
\dot z=e^{i\alpha}\overline{W'}\;.
\ee
Generically there will be no instantons for arbitrary value of $\alpha$. We will consider the case of the double well potential, with the superpotential already given after Eq.~(\ref{eq:lag}) 
\be
W(z)=\frac{z^3}{3}-a^2 z\;.
\ee
The BPS equations which give an (anti-)instanton solution are
\be
\dot z=\pm \overline{W'}\;.
\ee
This equation is solved by
\be
\label{zet}
z=\mp a\tanh(at)\;.
\ee

We will call the solution with the upper sign an \emph{instanton}, and the one with the lower sign an \emph{anti-instanton}.
The instanton solution breaks half of the supersymmetries (\ref{susyeqns}). In particular, an instanton background is invariant under SUSY with parameters $\bar\epsilon_1=\epsilon_2, \bar\epsilon_2=-\epsilon_1$, but under the remaining SUSY transformations with $\epsilon=\bar\epsilon_1=-\epsilon_2$ and $\tilde\epsilon=\bar\epsilon_2=\epsilon_1$, the fermionic fields become
\begin{align}
&\delta \chi_1=-2\sqrt2\dot z \tilde\epsilon\;,&&\delta\bar\chi_1=-2\sqrt{2}\dot{\bar z}\epsilon\;,\\
&\delta\chi_2=2\sqrt{2}\dot z\epsilon\;,&&\delta\bar\chi_2=-2\sqrt2\dot{\bar{z}}\tilde\epsilon\;.
\end{align}
The fermions depending on $\epsilon$ and $\tilde\epsilon$ can be, respectively, combined into  two-component spinors, 
 omitting the Grassmann factors of $\epsilon$, $\tilde\epsilon$:
 \begin{align}\label{zeromodes1}
&\xi=N\begin{pmatrix}\dot z\\
-\dot {\bar z}\end{pmatrix}   \;,&&\bar\xi=N\begin{pmatrix}\dot {{z}}\\
\dot {\bar{z}}\end{pmatrix} .
\end{align} where we introduced a  normalization factor $N$ (it is easily seen that $N^2 = 3/(8 a^3)$ for unit-normalized fermions).
The fermions $\xi$ and $\bar\xi$ are respective zeromodes of the Weyl operator $D$ and its hermitean conjugate  
\begin{align}\label{weyl}
&D=\partial_t+\begin{pmatrix}0&\overline{W''(z)}\\
{W''(z)}&0\end{pmatrix}\;,&&  D^\dagger=-\partial_t+\begin{pmatrix}0&\overline{W''(z)}\\
{W''(z)}&0\end{pmatrix}\;. 
\end{align}
Thus, an instanton always has two zeromodes of opposite chirality (in accordance with the index theorem, 
$\text{dimKer}D D^\dagger - \text{dimKer}D^\dagger D  =0$ for any background). This has important consequences in what follows, allowing zero modes to get lifted by perturbative effects. 

\subsection{Strategy and guide to calculation}

In this section we will calculate the two contributions to the instanton--anti-instanton amplitude $[I\bar I]$. The two contributions that need to be calculated are
\begin{itemize}
\item The fermion correlated amplitude $[I\bar I]_F$ (Top of Fig. \ref{fig:diagrams}),
\item The  Yukawa-scalar-exchange correlated amplitude $[I\bar I]_Y$ (Bottom of Fig. \ref{fig:diagrams}).
\end{itemize}

The most important part of $[I\bar I]_F$ amplitude calculation is  that  the instanton fermion zeromode is  lifted by the presence of the anti--instanton.  We therefore must carefully compute the lowest mode of the fermion operator in the instanton--anti-instanton background. The way we do this is by applying the standard degenerate perturbation theory. In short the lowest mode of the fermionic operator is proportional to the \emph{matrix element of the fermionic operator in the unperturbed zeromode basis} (see \eqref{eq:IbarIzeromode} below). This gives the non-trivial part of the result for $[I\bar I]_F$ given in \eqref{fexchange}.

To compute the scalar  correlated amplitude $[I\bar I]_Y$, we first find the zeromode lifting for arbitrary background field $y(t)$ in addition to an instanton. This background will lift the instanton zeromode to \eqref{overlap} below. The same is true for an anti-instanton in the background $y(t)$. Next we must integrate out the background $y(t)$ field, which requires us to use the propagator in the background of an instanton--anti-instanton (see \eqref{scalarexchange1}), which is well approximated by \eqref{ypropagator}. Armed with this knowledge, we are finally able to produce the result for $[I\bar I]_Y$ in \eqref{scalarexchangefin}.

\subsection{Fermion zeromode exchange}
Before we discuss this effect, we first  study  the lifting of the zero modes of an instanton due to the presence of an anti-instanton (or vice versa). In other words, we  consider a configuration with an instanton $I$ at time $t_1$ and an anti-instanton $\bar{I}$ occurring at time $t_2$. An approximation to this $I$-$\bar I$ configuration, valid at large separation $|t_2 - t_1| \gg 1/a$, is  
\be
\label{xbckgd}
x(t)=x_1(t)+x_2(t)+a
\ee
where
\be
x_1(t)=-a\tanh(a(t-t_1))\;,\qquad x_2(t)=a\tanh(a(t-t_2))\;,
\ee
with $t_2>t_1$.\footnote{For $t_2<t_1$ one must take $x=x_1+x_2-a$.}

We now use the Weyl operators $D$ and $D^\dagger$ of Eq.~(\ref{weyl}) to define the antihermitean Dirac operator\footnote{\label{diracfootnote}We define a Dirac operator as it has definite hermiticity properties and standard degenerate perturbation theory can  be used to compute the lifting of zero modes. The fermion part of (\ref{eq:lag}) is now $-{1\over 2} \Xi^T \slashed D \Xi$, with $\Xi^T = (\chi_1, \bar\chi_2, \bar\chi_1, \chi_2)$. Integrating out $\Xi$ gives then the Pfaffian of $\slashed D$, a fact  used in (\ref{pfaffian}). } 
\be
\slashed D=\begin{pmatrix}
0&D\\
-  D^\dagger&0
\end{pmatrix}.
\ee
In the $I$-$\bar{I}$ background, $\slashed D$ no longer has any zeromodes. But in the limit $|t_1-t_2|\rightarrow \infty$, the zeromodes of the instanton and the anti-instanton become exact. They are given by
\begin{align}
\label{zeromodes}
&\Psi_{1,2}=\begin{pmatrix}
0\\
\xi_{1,2}
\end{pmatrix}\;,&&\overline\Psi_{1,2}=\begin{pmatrix}
\bar\xi_{1,2}\\
0
\end{pmatrix}
\end{align}
where
\begin{align}
\label{xis}
&\xi_{1,2}=N\begin{pmatrix}
\dot x_{1,2}\\
\mp\dot x_{1,2}
\end{pmatrix}\;,&&\bar\xi_{1,2}=N\begin{pmatrix}
\dot x_{1,2}\\
\pm\dot x_{1,2}
\end{pmatrix}
\end{align}
where upper signs are for $I$, located at  $t_1$, as in (\ref{zeromodes1})  and lower signs for $\bar{I}$, located at   $t_2$ ($\bar\Psi$ denotes a separate spinor not to be confused with the complex conjugate to $\Psi$).
Making the separation finite will lift the zero eigenvalue.  To compute the lifted eigenvalue, 
we look for a solution
\be
\slashed D\Psi=i\varepsilon\Psi\;.
\ee
Using degenerate perturbation theory, it is straightforward but tedious to show that the eigenvalue $\varepsilon$ to leading exponential accuracy in $\tau$ is lifted to
\be\label{eq:IbarIzeromode}
\varepsilon\approx \pm \braket{\overline\Psi_2}{\slashed D}{\Psi_1}\approx  \pm 12 ae^{-2a\tau}\;.
\ee
Integrating out fermions we obtain the Pfaffian of $\slashed D$
\be
\label{pfaffian}
\text{Pf}\slashed D=\sqrt{\det \slashed D}=\varepsilon^2(12 a)^2e^{-4a\tau}\sqrt{\det{'}\slashed D}
\ee
where the prime signifies that zero modes have been excluded. Nonzero mode determinants are known to factorize at large separations $\tau$. Then we can write the $I\bar I$ fermion correlated contribution as
\be
\label{fexchange}
[I\bar I]_F=36\omega^2 e^{-2\omega\tau} e^{-2S_0-S_{int}} d\mu_I d\mu_{\bar I}
\ee
where $\omega=2a$, $d\mu_{I},d\mu_{\bar I}$ are the $I$ and $\bar I$ measures, including the translational moduli measures as well as the non-zeromode (factorized) determinants, $S_0$ is the action of the (anti-)instanton. Finally, $S_{int}$, the interaction action between the instanton and anti-instanton  at large separation (it can be easily derived or seen in, e.g.~\cite{ZinnJustin:1981dx})  between $I$ and $\bar I$ is given by
\be
\label{interaction}
S_{int}= - 12 S_0 \;e^{- 2 a \tau} = -\frac{32 a^3}{g}e^{-2a\tau}=-\frac{4 \omega^3}{g}e^{-\omega\tau}\;.
\ee
The contribution that we just computed---the lifting of the fermion zero mode in an $I$ background due to the presence of an $\bar I$ (or v.v.)---can be interpreted as due to the fermion exchange diagram on the top of Fig.~\ref{fig:diagrams}. Thus, Eq.~(\ref{fexchange}) gives the fermion-exchange $I$-$\bar{I}$ contribution to the vacuum energy. 

\subsection{Scalar exchange}
As already mentioned, there is another contribution to the $I$-$\bar I$ molecule. This contribution comes from the fact that the fermionic zero modes can be lifted by perturbing the instanton with a $\delta z=iy$ fluctuation. In other words, a    scalar $y$ fluctuation can lift the fermionic zero modes rendering the instanton contribution non-vanishing. In 
Fig.~\ref{fig:diagrams} (bottom), this amounts to soaking up the fermionic zero modes into the scalar via a Yukawa term. 

To that end consider the background field $x_1(t)$  and fluctuations
\be
z=x_1(t)+iy(t)\;.
\ee
where $y(t)$ is arbitrary, but small (so that it can be treated as a perturbation) and $x_1(t)=-a\tanh(at)$ is an $I$ background.\footnote{The fact that fluctuations around $x_1(t)$ in the Re($z$)=$x$ direction do not lift the zero modes  follows from the vanishing of the overlap integrals (\ref{overlap}) with $\tau^2$ replaced by $\tau^1$. }

The Weyl operator is $D=D_I+2y(t)\tau^2$, where $D_I=\partial_t+2x_1(t)\tau^1$ is the Weyl operator in the instanton background. In the same way as before, we compute the lowest Dirac eigenvalue by computing the matrix element of the Dirac operator (taken in the instanton plus $y$-fluctuation background) in the zero mode basis 
\be
\label{overlap}
\varepsilon= - i \int dt\;  \overline \Psi_1^T  \slashed D \Psi_1 = - i{2}\int dt\; \bar\xi^{\; T}  y\tau^2\xi = 4 N^2 \int dt\; \dot{x_1}(t)^2 \;y(t)= \frac{3 a}{2} \int dt\;  \frac{y(t)}{\cosh^4(at)}~,
\ee
where $\Psi_1$ are unit-normalized four-component spinors (\ref{zeromodes}) composed of the $\xi$, $\bar\xi$   zero modes from (\ref{zeromodes1}) (the value of $N$ is  given there)  and $x_1(t)$ is the instanton solution  (\ref{zet}).
In other words, we find that an instanton at position $t_1$ couples to the background $y$-field as 
\be\label{eq:instyuk}
[I]_y=\frac{3 a}{2} \int dt \frac{y(t)}{\cosh^4(a(t- t_1))} e^{-S_0} d\mu_I.
 \ee
One can interpret this result as follows: Formally, the fermion zero mode structure of an instanton is $ \sim e^{-S_0}  \chi_1 \chi_2 (t_1)  d\mu_I$ and the Yukawa term in the action  is $\int dt \bar \chi_1  \bar \chi_2 y $. The instanton amplitude is thus modified into \eqref{eq:instyuk} where the kernel is the square of the zero mode wave-function. 
Note that the support of the kernel is  $a|t-t_1| \lesssim 1 $, and thus, the modified instanton amplitude is roughly $[I]_y  \sim   y(t_1) e^{-S_0} d\mu_I$, 
where fermion zeromodes are converted into a scalar. However, we will need the exact kernel and expressions in order to show our main results. 
 Repeating the same for the anti-instanton, we find the same coupling of $y(t)$ to an anti-instanton at $t_{2}$. 
 Because the average  $\avg{y(t)}=0$, the single-instanton events do not contribute to the ground state energy.
 
 On the other hand, the  $I$-$\bar I$ scalar-correlated event may and does contribute to the ground state energy. 
  The  contribution  is
 \be
 \label{scalarexchange1}
[I\bar I]_Y=\frac{9 a^2}{4} \int dt\int dt' \frac{\avg{y(t)y(t')}}{\cosh^4(a(t-t_1))\cosh^4(a(t'-t_2))}e^{-2S_0-S_{int}}d\mu_I d\mu_{\bar I}~,
\ee
where $\avg{y(t)y(t')}$ is the scalar propagator in the $I$-$\bar I$ background. The other factors in (\ref{scalarexchange1})---measure, nonzero mode determinants, action---are the same as in the $[I\bar I]_F$ fermion-correlated event whose contribution is given in (\ref{fexchange}). Notice that (\ref{scalarexchange1}) can be equivalently viewed as due to two Yukawa-coupling insertions, taken in the $I$/$\bar I$ zeromode basis, and a scalar propagator from $I$ to $\bar I$---as pictorially shown in the bottom diagram of Fig.~\ref{fig:diagrams}.

\vspace{0.5cm}
\noindent{\bf $y$-propagator in the  $I$-$\bar I$ background:}
What remains is to find the $y$-propagator in the $I$-$\bar I$ background and compute the integral in (\ref{scalarexchange1}). To begin, note that 
 to quadratic order in $y$, we have the action  in the $I$-$\bar I$ background $x(t)$ of (\ref{xbckgd})
\be
S_y=\frac{1}{g}\int dt\;  y(-\partial_t^2+(2x^2+2a^2))y~,
\ee
so that
\be
\avg{y(t)y(t')}=\frac{g}{2}\frac{1}{-\partial_t^2+(2x^2+2a^2)}=\frac{g}{2}\;G(t,t'; t_1,t_2)\;,
\ee
where $G(t,t'; t_1,t_2)$ denotes the propagator in the $I$-$\bar I$ background.

The exact computation of $G(t,t'; t_1,t_2)$ is difficult, but for well-separated $I$ and $\bar I$ it can be approximated to sufficient accuracy by knowing the exact propagator in a single-instanton background.
For a single instanton  located at $t_0$, the $y$-propagator is
\begin{eqnarray}
G_{I}(t,t',t_0)&=&-\frac{1}{12 a}e^{-2a|t-t'|}(2\; \sign(t-t')+\tanh(a(t-t_0)) (-2\;\sign(t-t')+\tanh(a(t'-t_0))\nonumber \\&\equiv&g(t,t',t_0)\;G_0(t-t')~,
\end{eqnarray}
where we introduced the functions
\begin{align}
\label{gfunction}
&g(t,t';t_0)=-\frac{1}{3}\left\{2\;\sign(t-t')+\tanh[a(t-t_0)]\right\}\\\nonumber&\hspace{4cm}\times\left\{-2\;\sign(t-t')+\tanh[a(t'-t_0)]\right\}\;,\\\label{freeprop}&G_0=\frac{1}{4a}e^{-2a|t-t'|}~.
\end{align}
 This expression can be derived in many ways; an easy check is to verify that it obeys the appropriate equation with a delta-function source.
 
Notice that the $y$-propagator in an $I$ background $G_I$ is always proportional to $G_0$, the free propagator of the $y$-field (the same in either vacuum) and that  for  fixed sign$(t-t')$ the function $g$ is approximately constant except for $t$ or $t'$ near the instanton. 
Thus, a characteristic feature of $G_I$ is that it exhibits integer jumps (in units of $G_0$) whenever either $t$ or $t'$ cross $t_0$.    When the points $t$ and $t'$ are on the right of the instanton, and sufficiently far,  indeed, as expected on intuitive grounds, the $y$-propagator is just free propagator.  On the other hand, when the points $t\gg t_0$ and $t'\approx t_0$, 
the $y$-propagator is twice  free propagator. Finally, if $t\gg t_0$ and  $t' \ll t_0$, the $y$-propagator is enhanced by a factor of three with respect to the free propagator.  This effect, we believe, is tied up with the space being one dimensional, where the instanton eases the propagation of $y$-fluctuations compared to the vacuum $y$-fluctuations.

These features can be used to argue that for  a well-separated $I$-$\bar I$ background,  the $y$-propagator is  approximated with sufficient accuracy  by the product of the free propagator $G_0$ and the (identical) $g$-functions for an $I$ and $\bar I$:
\be
\label{ypropagator}
G(t,t';t_1,t_2)=g(t,t';t_1)\; g(t,t';t_2)\; G_0(t-t')\;.
\ee
with $G_0(t-t')$ is the free propagator \eqref{freeprop} and $g(t,t;t_i)$ is defined in \eqref{gfunction}.
The upshot is that we now have the desired expression for the $y$ propagator in the $|t_2-t_1|\gg 1/a$ $I$-$\bar I$ background (corrections to  (\ref{ypropagator}, \ref{ypropagator2}) can be seen to be of order $e^{-4 a |t_1 - t_2|}$, beyond our intended accuracy):
\be
\label{ypropagator2}
\avg{y(t)y(t')}=\frac{g}{8 a}e^{-2a|t-t'|}\;g(t,t';t_1)\;g(t,t';t_2)\;,
\ee 
Therefore,   \eqref{scalarexchange1} becomes
\be
[I\bar I]_Y=\frac{9 ag}{4\times 8} \int dt\int dt' \frac{e^{-2a|t-t'|} g(t,t';t_1)g(t,t';t_2) }{\cosh^4(a(t-t_1))\cosh^4(a(t'-t_2))}e^{-2S_0-S_{int}}d\mu_I d\mu_{\bar I}\;.
\ee
Since we only consider configurations for which $|t_2-t_1|\gg 1/a$, and since the fermion zeromode wavefunctions localize $t$ near $t_1$ and $t'$ near $t_2$, we may take the limit $|t'- t|\gg 1/a$. Then the expressions for $g$-functions (\ref{gfunction}) simplify
\begin{align}
&g(t,t';t_1)\approx \left(2-\tanh[a(t-t_1)]\right)\;,\\
&g(t,t';t_2)\approx \left(2+\tanh[a(t'-t_2)]\right)
\end{align}
where we assumed that $t'>t$. The amplitude then becomes
\begin{multline}
[I\bar I]_Y\approx \frac{9 ag}{4\times 8} \int_{-\infty}^\infty dt \frac{e^{2a t}(2-\tanh(a(t-t_1))}{\cosh^4(a(t-t_1))}\\\times\int_{-\infty}^\infty dt' \frac{e^{-2a t'}(2+\tanh(a(t'-t_2))}{\cosh^4(a(t'-t_2))}e^{-2S_0-S_{int}}d\mu_I d\mu_{\bar I}\;.
\end{multline}
One can easily do the integrals
\begin{align}
 \int_{-\infty}^\infty d(at) \frac{e^{2a t}(2-\tanh(a(t-t_1))}{\cosh^4(a(t-t_1))}= 4e^{2at_1}\;,\\
 \int_{-\infty}^\infty d(at') \frac{e^{-2a t'}(2+\tanh(a(t'-t_2))}{\cosh^4(a(t'-t_2))}=4e^{-2at_2} 
\end{align}
which gives
\be\label{scalarexchangefin}
[I\bar I]_Y=\frac{9 g}{2a} e^{-2a(t_2-t_1)}e^{-2S_0-S_{int}}=\frac{9g}{\omega}e^{-\omega\tau}e^{-2S_0-S_{int}}d\mu_{I}d\mu_{\bar I}
\ee
where $\tau=t_2-t_1$, and the interaction action $S_{int}$ at large separation is given in \eqref{interaction}. This is the contribution of the scalar-exchange induced correlated
event into the ground state energy.

\subsection{The magic of the thimble}
Putting all together,  the fermion and scalar correlated instanton-anti-instanton event, 
we arrive at our main result:
\begin{multline}
\label{iibarfinal}
[I\bar I]=[I\bar I]_F+[I\bar I]_Y=e^{- 2 S_0} \; \frac{1}{\omega}e^{\frac{4\omega^3}{g}e^{-\omega\tau}}\\\times\left(36\omega^3 \int_{\mathcal J_1}\;e^{-2\omega\tau}d\mu_{I}d\mu_{\bar I}+9g \int_{\mathcal J_2}\;e^{-\omega\tau}d\mu_{I}d\mu_{\bar I}\right)~,
\end{multline}
where $\mathcal J_1$ and $\mathcal J_2$ are the integration cycles on the corresponding thimbles in \eqref{eq:J1J2thimble}.
As promised, it has precisely   the form given in \eqref{eq:groundstate}. It is worthwhile repeating the main messages:
\begin{itemize} 
\item{Naive integration over the separation quasi-zero mode, viewing $\omega \tau \in \R^{+} $,  leads to erroneous positive  $[I\bar I]$ amplitude, or negative 
ground state energy, as the integrand is positive-definite on the naive integration cycle (see  Fig.\ref{fig:cycles}). This clearly  contradicts  to basic implication of supersymmetry algebra, the 
positive semi-definiteness of the ground state energy in a supersymmetric theory. 
 }
\item{The vacuum energy vanishes after integration over the appropriate Lefschetz thimbles,  ${\cal J}_1+   {\cal J}_2$, (see  Fig.\ref{fig:cycles}),  by the reasoning
 explained in Section~\ref{sec:formulation-2}. There is a relative phase, a counterpart of the hidden topological angle (HTA), between the   ${\cal J}_1$ and   ${\cal J}_2$ contribution.}
 \item{This provides concrete evidence, along with  Ref.~\cite{Behtash:2015kna,Behtash:2015,BDSSU},  that the proper framework to study multi-instanton amplitudes is the integration over the  Lefschetz thimbles. We believe that this results is universal and applies to general QFTs.}
 \end{itemize}
Notice that  (\ref{iibarfinal}) combines into a double total derivative 
\be \label{total}
[I\bar I]=e^{- 2 S_0} \; \frac{9g^2}{4\omega^6}\; \partial_{\tau}^2  e^{-S_{int}(\tau)} \; d\mu_Id\mu_{\bar I}\;.
\ee
The appearance of a total derivative at large $I$-$\bar I$ separation is a consequence of SUSY and has been observed long ago by Yung in 4D ${\cal{N}}=1$ SQCD \cite{Yung:1987zp}.
Taking (\ref{total}) literally and {\it assuming} its validity at all separations $\tau \in (0, +\infty)$, i.e.~along the entire ``streamline" \cite{Balitsky:1986qn}, one could argue that the integral over $\tau$ of (\ref{total}) has a piece at $\tau\rightarrow +\infty$ which clearly vanishes, and a piece at $\tau=0$, which is assumed to vanish, as an $I$ and $\bar I$ on top of each other are taken to represent the (zero) perturbative vacuum contribution in a SUSY theory.\footnote{This  is the argument from \cite{Yung:1987zp}. The essential difference is that there, because of the minimal amount of SUSY in 4D, the result is  a single total derivative w.r.t.~the quasi-zero mode. The contribution at infinity  gives the $I$-$\bar I$-induced potential, usually derived from an exact superpotential, on the moduli space.} 

What is remarkable is that the same result is obtained without any use of the supersymmetry constraint and without any assumptions about the streamline. The method  presented here is applicable to any system regardless of the supersymmetries. We stress that, as 
 opposed to the streamline, on the Lefschetz thimble  the separation $\tau$ between $I$ and $\bar I$ is never zero. As a result, the field configurations one integrates over are always distinct  from the perturbative vacuum. The saddle-point value (\ref{saddletau}) of $\tau$,  of order ${1 \over \omega} \log {\omega^3 \over g}$, gives the size of the $I$-$\bar I$ molecule. Thus, in the semiclassical $g \ll \omega^3$ limit  the  $\tau \gg 1/\omega$   approximation used throughout our derivation is  valid.\footnote{\label{footnotevalidity}The smallest  (by absolute value) separation between $I$ and $\bar I$ along the thimble is $\tau_{min} = {i \pi \over \omega}$. Strictly speaking, the use of the well-separated $I$-$\bar I$ configuration at such values of the separation is not   justified. However, it is easy to see from (\ref{iibarfinal}) that   the contribution to the integral from this small-$|\tau|$ region is  exponentially 
suppressed w.r.t.~the  $e^{- 2 S_0}$ accuracy of our second-order semiclassical approximation. }

\subsection{Remark on the BZJ-prescription}
Finally, a brief remark on the BZJ-prescription  \cite{ZinnJustin:1981dx,ZinnJustin:1982td}  is in order. According to BZJ, before integrating over  the quasi-zero mode separation, 
$\tau \in {\mathbb R^+}$, we first need to take $g\rightarrow -g$. Doing so, the $S_{int}$ part in the   instanton-anti-instanton interaction becomes repulsive,  while  scalar-exchange induced and fermi--zeromode exchange induced attractive  interactions remain unaltered. 
We can do both integrations there on $\tau \in {\mathbb R^+}$.  Then, we are supposed to reverse continuation back  to the physical theory, $-g \rightarrow e^{i \pi} (-g)$.   
 In principle, one may think that this should be equivalent to the integration over thimbles, because the reverse continuation may be viewed as the shift of the integration cycle 
$ {\mathbb R^+}$ to  ${\mathbb R^+}  + i\pi$.  But these are  not exactly the desired thimbles  ${\cal J}_1$ and   ${\cal J}_2$, rather only the positive halves of them, ${\rm Re} \;\tau \geq 0$. Is this good enough? The answer, in this theory, is ``no." To see this, let us perform the integrals in (\ref{eq:groundstate})  by first taking $g\rightarrow - g$ and integrating over $\tau$ from $0$ to $\infty$ (recalling that $S_0 = \omega^3/(3 g)$):
\begin{equation}
\label{eq:BZJthimble}
E_0(-g) \propto - e^{ 2 S_0} \int\limits_0^\infty d \tau e^{- {4 \omega^3\over g}e^{- \omega \tau}} \left( 4 \omega^3 e^{- 2 \omega \tau} - g e^{- \omega \tau}\right) =  {g \over \omega} \; e^{  2 S_0} \; e^{-  {4 \omega^3 \over g}} = +{g \over \omega} e^{  -{10 \omega^3 \over 3 g}} ~.
\end{equation}
Then, following BZJ, we continue back to positive $g$:
\begin{equation}
\label{eq:BZJthimble1}
E_0(g) \propto - {g \over \omega} e^{  +{10 \omega^3 \over 3 g}} ~.
\end{equation}
Thus, the BZJ prescription results in a contribution to the ground state energy that is {\it a.)} negative, in clash with unbroken supersymmetry,  and {\it b.)} exponentially large for physical values of $g$. Presumably, this exponentially growing contribution should be discarded (as was tacitly assumed in \cite{ZinnJustin:1981dx,ZinnJustin:1982td}), but the rationale for doing so does not clearly follow from the BZJ prescription. 
On the other hand, within the thimble integration, the contribution to the ground state energy vanishes, up to ${\cal{O}}(e^{- 4 S_0})$ subleading-order terms (see also Footnote \ref{footnotevalidity}). The role of thimbles for avoiding exponentially growing contributions was noted  in \cite{CDU}.
 
The above considerations force  us to view the integration over the Lefschetz thimbles as a rigorous  version of the BZJ-presciption.  Furthermore, thimbles geometrize the BZJ prescription. The semi-classical  method  instructs us that  the  integration over the separation quasi-zero mode must be done on the manifolds of 
complex gradient flows, and in our opinion, makes it  more intuitive.  (Despite the fact that it also forces us  to abandon the perspective that the separation between the instanton-anti-instanton for a correlated event is real.)

\section{Discussion and Conclusion}
\label{sec:discuss}

This is the curious incident of instantons and instanton--anti-instantons in the ${\cal N}=2$ supersymmetric QM.  Sometimes,  not the presence of something,  but rather the  absence  thereof, is an intriguing  phenomenon. The  story we described  here is such.   The absence of an interesting instanton-anti-instanton effect in the supersymmetric ${\cal N}=2$  QM, leads us to concrete 
conclusions about the nature  of the semi-classical method in QM and QFT.

Despite the fact that both fermion-exchange  induced and scalar-exchanged  induced instanton anti-instanton contributions to the ground state energy 
are: {\it i)} Naively, negative definite and {\it ii)} Formally, of different order in the coupling due to lifting of fermi zero modes by Yukawa couplings 
(with no hope of cancelling each other out), a different story develops when the integrations are performed on QZM Lefschetz thimbles. 
On the thimbles, the phases  (the hidden topological angles  \cite{Behtash:2015kna})  of these two contributions differ by a factor of $\pi$, 
\begin{align}
\arg {\cal J}_1 = \arg {\cal J}_2 + \pi
\label{HTA-2}
\end{align}
 Furthermore, the formally different order of the two contributions in the perturbative  coupling parameter $g$ is  compensated by the fact that fermion  exchange  and boson exchange induced attractions are of different order in separation.  Consequently, the contribution to ground state energy vanishes, as it must. 
 
 If the non-perturbative vacuum of the ${\cal N}=2$  theory is  described in terms of a dilute gas of fermion-correlated  $[I\bar I]_F$  and scalar-correlated  
 $[I\bar I]_Y$  two-events,  then, these are, in an Euclidean  description, excursions from one well to  the other and back.  The reason that the two contribution do not give a net contribution to the ground state energy is the relative hidden topological angle \eqref{HTA-2} associated with these two kinds of tunnelling events. 
 
 The procedure of using thimbles  implicitly omits the contribution of the perturbative vacuum---to which the instanton--anti-instanton contribution is continuously connected. It is, in principle, applicable to any theory where the semiclassical expansion is justified. Although the procedure does not classify \emph{all} the saddles which may contribute to the various observables, it appears to be a necessary ingredient of the semiclassical expansion.

We also  remark on a QFT in which a similar effect may be operative. It is known  that in ${\cal N}=1$ SYM on small $\R^3 \times \S^1$,  both magnetic bions and neutral bions are present, as can be deduced either by using the superpotential, the BZJ-prescription,  or the method of this paper. But in ${\cal N}=2$ SYM in the same small $\R^3 \times \S^1$ regime (see \cite{Poppitz:2011wy} where the question was raised) neither contribution should be there, as can be seen by resorting to supersymmetry---monopole-instantons do not induce a superpotential, because they have four fermi zero modes. Similarly, the neutral bions do not form in $\mathcal N=1$ SQCD with massless flavors on $\R^3\times\S^1$ \cite{Poppitz:2013zqa}. We believe that a mechanism similar to the one described in this work also explains the absence of bosonic potential in these theories.

Finally, we remark on another QFT application. In many QCD or SQCD type theories on $\R^4$, the small instanton contributions are  calculable, even though large-instantons may be incalculable. In this context, it is well-known that the interaction between two instantons do not only depend on the 
separation, but also on orientational quasi-zero mode.  Depending on  the relative orientation,  the interaction between two-instantons 
may be both attractive and repulsive.   It may be worthwhile to look at this type of system by using appropriate thimbles. 

 \acknowledgments
 
We would like to thank  G\"okce Ba{\c s}ar, Aleksey Cherman, Marcos Crichigno, Daniele Dorigoni, Gerald Dunne,  Alyosha Yung, for useful discussions. This work was supported in part by a DOE grant DE-SC0013036. EP thanks North Carolina State University for hospitality during work on this paper and acknowledges support by an NSERC Discovery Grant.

\bibliography{bibliography}
\bibliographystyle{JHEP}

\end{document}